# Modified Hubble law, the time-varying Hubble parameter and the problem of dark energy

Jian-Miin Liu*
Department of Physics, Nanjing University
Nanjing, The People's Republic of China
* On leave. E-mail: liu@phys.uri.edu

In the framework of the solvable model of cosmology constructed in the Earth-related coordinate system, we derive the modified Hubble law. This law carries the slowly time-varying Hubble parameter. The modified Hubble law eliminates the need for dark energy.

Recently, based on Einstein's theory of gravitation and two assumptions of the cosmological principle and the perfect fluid for the matter of the Universe, in the so-called Z-approximation that the most probable velocity magnitude of the perfect fluid vanishes, we constructed a model of cosmology in the Earth-related coordinate system [1,2]. Here the cosmological principle states that the Universe is spatially homogeneous and isotropic on large scale with respect to the Earth-related coordinate system. This principle has its foundation in the cosmological measurements. With respect to the Earth-related coordinate system, the Hubble law of cosmological red shifts is direction-independent and the spectrum of cosmic microwave background is almost isotropic with a deviation at the level of one part in $10^5$ (The data made by satellites of the Earth, like COBE, during a relatively long period can be considered relative to the Earth-related coordinate system). The model is solvable. Its solution consists of three parts: the value for the cosmological constant, $\Lambda = 0$, the equation of state for the perfect fluid, $\rho + p = 0$, and the line element for space-time of the Universe [1,2],

$$ds^2 = (dx^0)^2 - \exp[2\sqrt{\frac{8\pi\rho G}{3}}t]\{dr^2 + r^2 d\theta^2 + r^2 \sin^2\theta d\phi^2\}, \qquad (1)$$

where $\{x^0, r, \theta, \phi\}$, $x^0 = ct$, denotes the Earth-related coordinate system in which the Earth is always at rest, $c$ is the speed of light in vacuum, G is the Newtonian gravitational constant, $\rho$ and $p$ are respectively the mass-density and pressure of the perfect fluid, and $t$ is relative to the SC-moment at which the space-time line element of the Universe is the Minkowskian.

The space-time line element of the Universe evolves with time. This evolution causes the red-shift phenomena of light signals coming from distant galaxies. Suppose a light signal of frequency $v_1$ is emitted at time $t_1$ on a galaxy located at distance $r$ from the Earth. This signal is later detected by us on the Earth at time $t$ to have frequency $v$. It is then compared to another light signal of the same type, emitted at time $t$ on the Earth, whose frequency is $v_0$. Experimental measurements indicate a red shift of the light signal emitted on the galaxy,

$$z \equiv \frac{v_0 - v}{v} > 0.$$

The emissions of two light signals, at $t_1$ on the galaxy and at $t$ on the Earth, are both of physical process under control of some stronger interaction other than gravity, occurring in a localized region of space.

They are beyond the physics of the gravitational field of the Universe. It is natural to recognize $v_0 = v_1$. The red-shift is thus

$$z = \frac{v_1 - v}{v} > 0. \tag{2}$$

The equation of motion for light propagation can be obtained by setting $ds^2 = 0$ in Eq.(1). For purely radial propagation, it is

$$\pm \exp[-\sqrt{\frac{8\pi\rho G}{3}}t]cdt = dr. \tag{3}$$

If the galaxy is motionless and the first wavecrest of the light signal is emitted at time $t_1$ on this galaxy and detected at time $t$ on the Earth, and if the next wavecrest is emitted at time $t_1 + \delta t_1$ on this galaxy and detected at time $t + \delta t$ on the Earth, then in accordance with Eq.(3), we have

$$-\int_r^0 dr = \int_{t_1}^t \exp(-\sqrt{8\pi\rho G/3}t)cdt = \int_{t_1+\delta t_1}^{t+\delta t} \exp(-\sqrt{8\pi\rho G/3}t)cdt, \tag{4}$$

because distance $r$ is fixed. We rewrite the second equality in Eq.(4) as

$$\int_{t_1}^{t_1+\delta t_1} \exp(-\sqrt{8\pi\rho G/3}t)cdt = \int_t^{t+\delta t} \exp(-\sqrt{8\pi\rho G/3}t)cdt,$$

and integrate it for

$$\exp(-\sqrt{8\pi\rho G/3}t_1)\delta t_1 = \exp(-\sqrt{8\pi\rho G/3}t)\delta t$$

when both $\delta t_1$ and $\delta t$ are small enough. The last equation implies

$$\frac{v_1}{v} = \exp[\sqrt{8\pi\rho G/3}(t - t_1)]. \tag{5}$$

On the other hand, from the first equality in Eq.(4) we have

$$\frac{\sqrt{8\pi\rho G/3}}{c} r \exp(\sqrt{8\pi\rho G/3}t) = \exp[\sqrt{8\pi\rho G/3}(t - t_1)] - 1. \tag{6}$$

Using Eqs.(5) and (6) in Eq.(2) gives rise to

$$cz = H(t)r, \tag{7a}$$

$$H(t) = \sqrt{\frac{8\pi\rho G}{3}} \exp(\sqrt{\frac{8\pi\rho G}{3}}t), \tag{7b}$$

as the formula of cosmological red shifts. At any fixed detection time, it is a linear relation that exists between the cosmological red shifts of light signals coming from distant galaxies and the distances of these galaxies from the Earth. The proportional coefficient of the linear relation, $H(t)$, varies time. It is always positive and increases slowly as time runs from the past to the future, though exponentially, because $(8\pi\rho G/3)^{1/2}$ is small. The increasing rate is $dH(t)/dt = (8\pi\rho G/3)\exp(\sqrt{8\pi\rho G/3}\,t)$. Comparing the formula to the Hubble law [3]: $cz = Hr$, where $H$ is the Hubble constant, we call the formula the modified Hubble law and $H(t)$ the time-varying Hubble parameter. The nowadays value of $H(t)$ equals the Hubble constant.

There is a problem, called the "missing mass" or dark energy problem, in the standard model of cosmology [4-12]. The measured mass-density of the gravity-interacting matter (normal and dark) of the Universe is too small to fit in with two other measurement results simultaneously: The Universe is spatially flat [13,14] and the Hubble constant $H$ falls between 70 and 72 km/sec/Mpc [15,16]. In the framework of the standard model, the two measurement results require $H = \sqrt{8\pi\rho G/3}$ or $\rho = 3H^2/8\pi G$, where $H = 70 \sim 72$ km/sec/Mpc. But the measured mass-density is

$$\rho \approx 0.3(3H^2/8\pi G). \tag{8}$$

A kind of energy, named the dark energy, is needed for making up around 70 percent of all energy. This dark energy is mysterious. So far no evidence has indicated its existence, even indirectly.

The problem of dark energy does not exist in our model of cosmology. In view of Eq.(7b), the nowadays value of the time-varying Hubble parameter is

$$H(t^*) = H = \sqrt{8\pi\rho G/3}\exp(\sqrt{8\pi\rho G/3}\,t^*), \tag{9}$$

not $\sqrt{8\pi\rho G/3}$, where $t^*$ is the nowadays cosmological year relative to the SC-moment. Putting Eq.(8) and the value of $H$ into Eq.(9), we can find

$$t^* = H^{-1}\sqrt{10/3}\,\ell n\sqrt{10/3} \approx +15.1 \text{ billion years.} \tag{10}$$

The plus sign here means "after the SC-moment". 15.1 billion years is a long period making the factor $\exp[\sqrt{8\pi\rho G/3}\,t^*]$ quite significant. The "missing mass" problem in the standard model of cosmology is really due to missing the factor $\exp[\sqrt{8\pi\rho G/3}\,t^*]$ in Eq.(9). Around 15.1 billion years ago, at the SC-moment, the time-varying Hubble parameter was $H(0) = \sqrt{8\pi\rho G/3} = \sqrt{3/10}\,H(t^*)$, a little more than half of its nowadays value.

In summary, the modified Hubble law can be derived in the framework of the solvable model of cosmology constructed in the Earth-related coordinate system. This law carries the slowly time-varying Hubble parameter. The modified Hubble law eliminates the need for dark energy.

**Acknowledgment**

The author greatly appreciates the teachings of Prof. Wo-Te Shen. The author thanks to Dr. J. Conway for his supports and helps.

**References**

[1]    Jian-Miin Liu, physics/0505035
[2]    Jian-Miin Liu, physics/0506164
[3]    E. Hubble, Proc. Natl. Acad. Sci., 15, 168 (1929)
[4]    A. Friedmann, Z. Phys., 10, 377 (1922)
[5]    G. Lemaitre, Ann. Soc. Sci. (Bruxelles), 47A, 49 (1927)
[6]    H. P. Robertson, Appl. Phys. J., 82, 284 (1935); 83, 187 (1936); 83, 257 (1936)
[7]    A. G. Walker, Proc. Lond. Math. Soc., 42, 90 (1936)
[8]    S. Weinberg, Gravitation and Cosmology, Wiley & Sons (New York, 1972)
[9]    S. Dodelson, Modern Cosmology, Academic Press (New York, 2003)
[10]   M. Trodden and S. M. Carroll, astro-ph/0401547
[11]   J. Lesgourgues, astro-ph/0409426
[12]   J. Garcia-Bellido, astro-ph/0502139
[13]   P. De Bernardis et al, astro-ph/0004404
[14]   J. R. Bond et al, astro-ph/0011378
[15]   J. R. Mould et al, ApJ., 529, 698 (2000)
[16]   W. L. Freedmann et al, ApJ., 553, 47 (2001)